\documentclass[aps,prl,longbibliography,showpacs,notitlepage,twocolumn,superscriptaddress,nofootinbib,preprintnumbers]{revtex4-2}
\usepackage{bbm}
\usepackage{mathrsfs}
\usepackage{epsfig}
\usepackage{soul,xcolor}
\usepackage{graphicx}
\usepackage{amsfonts}
\usepackage{amsthm}
\usepackage[figuresright]{rotating}
\usepackage{amssymb}
\usepackage{amsmath}
\usepackage{cancel}
\usepackage{dcolumn}
\usepackage{physics}
\usepackage{bm}
\usepackage{physics}
\usepackage{verbatim}
\usepackage{braket}
\usepackage[normalem]{ulem}
\usepackage[ruled,vlined,linesnumbered]{algorithm2e}
\usepackage{setspace}
\usepackage[colorlinks,linkcolor=blue,anchorcolor=blue,citecolor=blue,urlcolor=blue]{hyperref}

\newtheorem{theorem}{Theorem}

\begin{document}

\title{\scalebox{0.99}[1]{High-Accuracy Neural-Network Quantum States via Randomized Real-Time Dynamics}}

\author{John M. Martyn}
\affiliation{Physical and Computational Sciences, Pacific Northwest National Laboratory, Richland, WA 99354, USA}
\affiliation{Harvard Quantum Initiative, Harvard University, Cambridge, MA 02138, USA}
\affiliation{The NSF AI Institute for Artificial Intelligence and Fundamental Interactions, USA}

\author{Rod Rofougaran}
\affiliation{Physical and Computational Sciences, Pacific Northwest National Laboratory, Richland, WA 99354, USA}

\author{Samuel Stein}
\affiliation{Physical and Computational Sciences, Pacific Northwest National Laboratory, Richland, WA 99354, USA}

\author{Di Luo}
\affiliation{Department of Physics, Tsinghua University, Beijing 100084, China}
\affiliation{Department of Electrical and Computer Engineering, University of California, Los Angeles, CA 90095, USA}

\begin{abstract} 
  Neural-network quantum states (NQSs) exhibit a disparity in converging to the ground state during energy minimization: while the energy converges \textit{quadratically} with the error in the variational wave function, other observables converge only \textit{linearly}, which limits the accuracy of physical predictions. 
  Here we show how to close this gap by introducing a new algorithm that improves the accuracy of NQSs. Our method, termed \textit{dynamical averaging}, estimates expectation values by averaging over a state (or observable) evolved in real-time. Using insights from quantum information, we prove an up-to-quadratic suppression of error in arbitrary ground state observables, thus matching the fast convergence of the energy. Importantly, this improved accuracy requires neither further energy minimization nor increased model complexity, but rather exploits the effect of randomization on quantum states. 
  We demonstrate dynamical averaging on non-integrable spin models, where it accelerates convergence and substantially reduces errors in observables. 
\end{abstract}

\maketitle

\textit{Introduction}---
A contemporary line of work has leveraged machine learning to tackle quantum many-body problems~\cite{Carleo_2019_Machine}. This is epitomized by neural-network quantum states (NQSs)~\cite{Carleo_2017_Solving}, a class of powerful variational ans\"atze that parameterize quantum states by neural networks. Supported by the rich expressivity of neural networks~\cite{Gao_2017, Sharir_2022}, NQSs can accurately model complex quantum systems, including ground states, real-time dynamics, field theories, and quantum materials~\cite{Carrasquilla_2021, Gutierrez_2022, Sinibaldi_2023, Martyn_2023, luo2023pairing, Foster_2026}. As such, these methods hold broad applications in chemistry, condensed matter physics, and materials science~\cite{lange2024architectures, medvidovic2024neural, Foulkes_2001, hermann2023ab}.

Despite their success, NQSs still incur a small, but non-negligible, systematic error in approximating a target state, which limits their ability to accurately estimate quantities of interest. For example, in approximating a ground state, this error can affect the accuracy of relevant observables like correlation functions, electron densities, and dipole moments~\cite{Pfau_2020, Hermann_2020, Pfau_2024}. In fact, the error suffered by an arbitrary ground state observable can be significantly larger than the error in the energy, creating a disparity that limits physical predictions~\cite{Assaraf_2003, zhang2020variational, zhang2025improved}.

Conventional wisdom in machine learning suggests that reducing this systematic error requires additional optimization or a more expressive model. Remarkably, neither is necessary for NQSs: in this work, we introduce \textit{dynamical averaging} as a quantum-inspired algorithm that provably reduces systematic error through real-time evolution.
Crucially, this does not require further energy minimization nor increased model complexity.

In more detail, we make three main contributions in this work. First, (1) we draw upon results from quantum information to prove that randomly evolving an NQS in real-time provides an up-to-quadratic suppression of systematic errors (i.e. $\epsilon \rightarrow \mathcal{O}(\epsilon^2)$). 
Then, (2) we develop dynamical averaging as an algorithm that realizes this error suppression by averaging over real-time dynamics. We formulate this in both the Schr\"odinger and Heisenberg pictures, where either the NQS or target observable evolves in time, respectively. In the latter, we also develop an approximation that avoids explicit real-time evolution by adding a correction to the standard observable estimate. 
Lastly, (3) we showcase dynamical averaging on non-integrable spin models in 1D and 2D, where it achieves faster convergence and significantly reduces errors in correlation functions and magnetizations. 

Collectively, these contributions improve the accuracy of ground state observables, bringing their convergence closer to that of the energy and addressing the aforementioned disparity. 
These findings also reveal the surprising utility of time evolution as a tool for optimization, suggesting that real-time dynamics may become an important ingredient in future optimization algorithms.

\vspace{6pt}
\emph{Background}---
NQSs parameterize a state vector $|\psi_\theta\rangle$ by a set of parameters $\theta$ and minimize its energy to approximate the ground state according to the variational principle. The energy is estimated using variational Monte Carlo (VMC)~\cite{Foulkes_2001, becca2017quantum, Sorella_2005, neuscamman2012optimizing}, which expresses it as a classical expectation value over configurations $x$:
\begin{equation}\label{eq:energ_VMC}
    \langle \psi_\theta | H |\psi_\theta \rangle = \mathop{\mathbb{E}}_{x \sim |\psi_\theta (x)|^2} \big[ E_{\text{loc}}(x) \big] . 
\end{equation}
Here $\psi_\theta(x) = \langle x |\psi_\theta\rangle$ is the NQS in a basis $|x\rangle$, and $E_{\text{loc}}(x) = \tfrac{\langle x | H |\psi_\theta \rangle}{\langle x | \psi_\theta \rangle} $ is the \textit{local energy}. Thus, the energy can be estimated by sampling $|\psi_\theta(x)|^2$ and evaluating the mean of $E_{\text{loc}}(x)$. Analogous expressions hold for other observables, e.g. $\langle \psi_\theta |O| \psi_\theta \rangle$, upon replacing $E_{\text{loc}}(x)$ with the appropriate local observable $O_{\text{loc}}(x) = \tfrac{\langle x | O |\psi_\theta \rangle}{\langle x | \psi_\theta \rangle}$. This approach is also used to evaluate the gradient of energy with respect to $\theta$ and perform optimization.

\begin{figure}
    \centering
    \includegraphics[width=0.8\columnwidth]{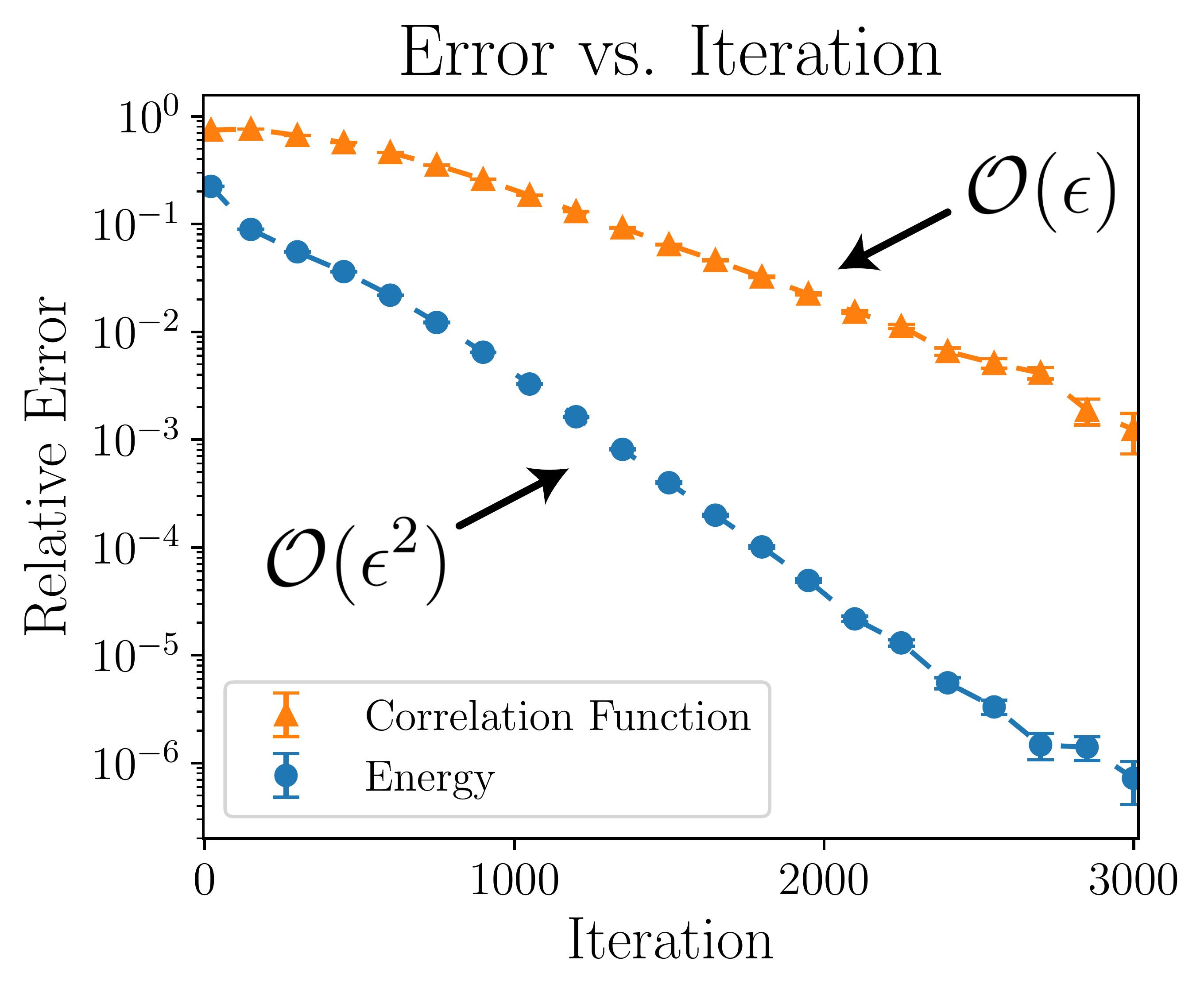}
    \caption{Application of an NQS to the mixed field Ising model on $N=12$ spins (see Experiments section below for details). We show the relative error in the energy and a correlation function at the center of the chain $\langle \sigma^x_{N/2} \sigma^x_{N/2+1} \rangle$.
    }
    \label{fig:Error_vs_Iteration}
\end{figure}

\vspace{6pt}
\emph{Motivation}---
Our work is motivated by a simple observation: when optimizing an NQS to approximate a ground state, the energy is often far more accurate than other observables. 
To illustrate this, Fig.~\ref{fig:Error_vs_Iteration} depicts the energy minimization process of an NQS applied to an Ising model. Here, we see that the error in energy decays faster than the error in a correlation function and reaches values a few orders of magnitude smaller. This discrepancy is notable because the energy does not entirely characterize the ground state; other observables like correlation functions and magnetization are often of equal physical interest.

This phenomenon is generic in NQSs (and other variational methods, like tensor networks) and stems from the ground state being an eigenstate of the Hamiltonian. Specifically, the energy converges \textit{quadratically} with the error in the wave function, while other observables only converge \textit{linearly}. To see this, consider an NQS that approximates the ground state as 
\begin{equation}\label{eq:NQS_GS_approx}
    | \psi_\theta\rangle = \sqrt{1-\epsilon^2}|E_0\rangle + \epsilon |E_\perp\rangle , 
\end{equation}
where $|E_0\rangle$ is the ground state, $|E_\perp\rangle$ is a linear combination of  excited states, and $\epsilon \ll 1$ characterizes the error.\footnote{Any ground state approximation may be expressed in this form. We can always take $\epsilon \in \mathbb{R}$ by absorbing a phase into $|E_\perp\rangle$.} 
Then, the relative error in an observable $O$ is:
\begin{equation}\label{eq:relative_error}
\begin{aligned}
    &\Bigg| \frac{ \langle \psi_\theta | O | \psi_\theta \rangle - \langle E_0 | O | E_0 \rangle }{ \langle E_0 | O | E_0 \rangle } \Bigg| = \\
    & \ \ \Bigg| 2 \epsilon \sqrt{1-\epsilon^2} \frac{\text{Re}\big( \langle E_0 | O | E_\perp \rangle \big)}{\langle E_0 | O | E_0 \rangle} + \epsilon^2 \bigg(\frac{\langle E_\perp | O | E_\perp \rangle }{ \langle E_0 | O | E_0 \rangle }  -1\bigg) \Bigg|.
\end{aligned}
\end{equation}
For an arbitrary observable, where $\langle E_0 | O |E_\perp\rangle \neq 0$ in general, the relative error scales as $\mathcal{O}(\epsilon)$. In contrast, for the energy ($O=H$), orthogonality of the eigenstates dictates that $\langle E_0 | H |E_\perp\rangle = 0$, which reduces the relative error to $\mathcal{O}(\epsilon^2)$. 

This observation can be formalized through a metric from quantum information known as the \textit{trace distance}, which measures the deviation between two density matrices $\rho$ and $\sigma$: $d_{\text{tr}} (\rho,\sigma) = \frac{1}{2} \| \rho - \sigma \|_1 $, where $\| \cdot \|_1$ denotes the Schatten 1-norm (sum of singular values).
The trace distance is especially useful as it bounds the difference in expectation values as~\cite{wilde2013quantum}:
\begin{equation}\label{eq:trace_dist_bound}
    | \tr(\rho O) - \tr(\sigma O) | \leq \| O \| \| \rho - \sigma \|_1  = 2 \| O \| d_{\text{tr}} (\rho,\sigma) ,
\end{equation} 
which is tight for arbitrary observables $O$. For the NQS in Eq.~\eqref{eq:NQS_GS_approx}, a straightforward calculation shows that its trace distance with the ground state is $d_{\text{tr}}\big( | \psi_\theta \rangle \langle \psi_\theta| , |  E_0 \rangle \langle E_0 | \big) = \mathcal{O}(\epsilon)$, which implies that $|\psi_\theta\rangle$ generally incurs $\mathcal{O}(\epsilon)$ errors in observables. However, because Eq.~\eqref{eq:trace_dist_bound} is only an upper bound, certain observables can attain smaller errors; as we saw above, the energy enjoys $\mathcal{O}(\epsilon^2)$ error scaling.

At first glance, this discrepancy might appear fundamental because the energy is being minimized and may naturally converge faster than other observables. 
However, when applying machine learning to physics, we can often surpass naive expectations by exploiting physical insights, like symmetries and equations of motion. This naturally raises the question: \textit{can we use physical insights to estimate observables with the same accuracy as the energy?} We answer yes, showing how randomized real-time dynamics reduces the trace distance to $\mathcal{O}(\epsilon^2)$, thus improving the accuracy of arbitrary observables.

\vspace{6pt}
\emph{Improving Accuracy with Real-Time Evolution}---Recent work in quantum information has shown that randomization can help in approximating unitary transformations~\cite{hastings2016turning, Campbell_2017, Campbell_2019}. Specifically, randomizing over many approximate transformations can outperform any single deterministic approximation of the same cost, achieving up to quadratically smaller error. 

Here we repurpose this concept to quadratically suppress errors in NQSs and resolve the disparity described above. In this setting, the analogous approach is to randomize over an ensemble of NQSs, which can approximate the ground state more accurately than a single NQS. Operationally, this amounts to taking a NQS, $|\psi_\theta\rangle$, and promoting it to a corresponding mixed state, which we denote by $\rho_\theta$. In doing so however, we will not seek to optimize over NQS representations of such mixed states, which could be costly. Rather, we introduce a simple procedure that maps $|\psi_\theta\rangle$ to $\rho_\theta$.

The key idea of our procedure is to randomly apply operations that behave differently on the eigenstates of $H$, enabling us to selectively suppress contributions from excited states. We realize this by evolving $|\psi_\theta \rangle$ in real-time for random durations, thus promoting it to the following mixed state:
\begin{equation}\label{eq:rho}
    \rho_\theta = \int dt\ p(t) e^{-iHt} |\psi_\theta \rangle \langle \psi_\theta | e^{iHt} , 
\end{equation}
where $p(t)$ is a probability distribution that we will specify later. Intuitively, this procedure imparts a random (relative) phase on the excited state components in $|\psi_\theta\rangle$, which suppresses their contributions.
Equivalently, this decoheres the off-diagonal elements of the density matrix $|\psi_\theta\rangle \langle \psi_\theta |$ in the energy eigenbasis, yielding a more accurate approximation of the ground state. We make this statement rigorous as follows:
\begin{theorem}\label{thm:main_theorem}
    Suppose an NQS $|\psi_\theta \rangle$ approximates a ground state $| E_0\rangle$ with trace distance error $d_{\text{tr}} \big( |\psi_\theta\rangle \langle \psi_\theta | , \ |E_0\rangle \langle E_0| \big) \leq \epsilon$. Then, the mixed state $\rho_\theta$ (defined in Eq.~\eqref{eq:rho}) suffers error 
    \begin{equation}\label{eq:RC_VMC_trace_dist}
        d_{\text{tr}}\big( \rho_\theta, \ |E_0\rangle \langle E_0 | \big) \leq \mathcal{O}(\epsilon^2) + \mathcal{O}\big( \epsilon\cdot \max_{n\geq 1} |\varphi(\Delta_n)| \big), 
    \end{equation}
    where $\Delta_n = E_n - E_0$ are the excitation energies, and $\varphi(\omega) = \int dt \ p(t) e^{i \omega t}$ is the characteristic function of $p(t)$. 
\end{theorem}

We include the proof of this theorem in the Supplementary Material. 
This result shows that randomizing over real-time evolution yields a more accurate ground state approximation. If the characteristic function is sufficiently small, i.e. $ |\varphi(\Delta_n)| = \mathcal{O}(\epsilon)$, then the trace distance is suppressed to $\mathcal{O}(\epsilon^2)$. Otherwise, the trace distance is bounded by $\mathcal{O}(\epsilon \cdot |\varphi(\Delta_n)|)$, which is still an improvement because $|\varphi(\omega)| \leq 1$ for all $\omega$. Therefore, $\rho_\theta$ is able to achieve an up-to-quadratic reduction in trace distance, which by Eq.~\eqref{eq:trace_dist_bound} translates to an equivalent reduction in the error of an arbitrary observable. This quadratic suppression is optimal in general; any stronger error suppression would imply a reduction in energy, which is not possible because real-time evolution conserves energy.

A remaining question is how to select a distribution $p(t)$ with a sufficiently small characteristic function. In practice, $\mathop{\max}_{n\geq 1} |\varphi(\Delta_n)|$ often evaluates to $|\varphi(\Delta_1)|$, and so bounding this quantity generally requires either a lower bound on the spectral gap $\Delta_1 = E_1-E_0$, or the ability to adaptively tune the distribution. For instance, if $p(t)$ is chosen to be a zero-mean Gaussian with variance $\sigma^2$, then $\varphi(\Delta_1) = e^{-\frac{1}{2}\sigma^2\Delta_1^2}$, and setting $\sigma^2 \geq \tfrac{2}{\Delta_1^2} \log(1/\epsilon)$ ensures $\varphi(\Delta_1) = \mathcal{O}(\epsilon)$. 
However, if $\Delta_1 $ is unknown, one can successively increase $\sigma^2$ until $\varphi(\Delta_1) = \mathcal{O}(\epsilon)$ and no further significant improvement in accuracy is observed. 
Alternatively, our recent work Ref.~\cite{arulandu2026trading} establishes a result analogous to Theorem~\ref{thm:main_theorem} for ground state preparation on quantum computers and derives optimal distributions for $p(t)$ given a lower bound on the spectral gap. These distributions can also be applied here.

Overall, replacing $|\psi_\theta\rangle $ with $\rho_\theta$ provides more accurate ground state observables. Realizing this in practice amounts to averaging over real-time evolution, applied to either the NQS or a target observable. As such, we call this algorithm \textit{dynamical averaging}, and we will use it to improve estimates of ground state observables below.

We remark that Theorem~\ref{thm:main_theorem} applies when an NQS is close to the ground state in trace distance, as occurs in spin chains of moderate size or molecules in a compact basis. However, this requirement can become difficult to satisfy in the thermodynamic limit due to the orthogonality catastrophe~\cite{Anderson_1967, Fogarty_2020, Erakovic_2025}. While this poses challenges for conventional ans\"atze like configuration interaction, recent studies suggest that NQSs may exhibit more favorable scaling with system size, potentially mitigating these challenges~\cite{viteritti2026approaching}.
In addition, as we show in the Supplementary Material, even in the thermodynamic limit dynamical averaging resolves a disparity between the energy and arbitrary observables. Specifically, the error in energy scales with only the diagonal elements of the state $|\psi_\theta\rangle \langle \psi_\theta|$ in the energy eigenbasis, whereas the error in an arbitrary observable receives contributions from both the diagonal and off-diagonal elements. Promoting $|\psi_\theta\rangle$ to $\rho_\theta$ then suppresses its off-diagonal elements by a factor of $\varphi(E_n - E_{n'})$. Under strong suppression, only the diagonal elements contribute appreciably to the error in an observable, recovering the same scaling as the energy.

\vspace{6pt}
\emph{Dynamical Averaging in the Schr\"odinger Picture}---
A direct way to simulate $\rho_\theta$ is through time evolution in the Schr\"odinger picture. 
In this context, we can use VMC to express an observable as a double expectation value:
\begin{equation}\label{eq:dyn_avg_eq}
\begin{aligned}
    \tr(O\rho_\theta) &= \int dt \  p(t) \langle \psi_\theta| e^{iHt} O e^{-iHt} | \psi_\theta\rangle \\
    &= \mathop{\mathbb{E}}_{t \sim p(t)} \ \bigg[ \mathop{\mathbb{E}}_{x \sim |\psi_\theta(x,t)|^2} \Big[ O_{\text{loc}}(x, t) \Big] \bigg]. 
\end{aligned}
\end{equation}
Here $\psi_\theta(x,t) = \langle x |e^{-iHt} |\psi_\theta\rangle$ is the time-evolved NQS in a basis $|x\rangle$, and $O_{\text{loc}}(x , t) = \langle x | O e^{-iHt} |\psi_\theta\rangle / \langle x | e^{-iHt} | \psi_\theta\rangle $ is the time-evolved local observable. In contrast to standard VMC which samples configurations from $|\psi_\theta(x)|^2$, Eq.~\eqref{eq:dyn_avg_eq} suggests sampling from time-evolved states $|\psi_\theta(x,t)|^2$ and averaging over times drawn from $p(t)$. 

A convenient way to realize this double expectation value is to evolve $|\psi_\theta(t)\rangle = e^{-iHt}|\psi_\theta\rangle$ in real-time, and collect expectation values $\langle \psi_\theta(t) | O | \psi_\theta(t) \rangle = \mathop{\mathbb{E}}_{x \sim |\psi_\theta(x,t)|^2} \big[ O_{\text{loc}}(x, t) \big]$ during evolution. These expectation values can then be aggregated together according to Eq.~\eqref{eq:dyn_avg_eq}. 
In practice, when time evolution is implemented as a sequence of time steps, we realize dynamical averaging by evolving $|\psi_\theta (t) \rangle$ in time and estimating the target observable at  a collection of time steps. 
These expectation values are then weighted by $p(t)$ and summed to estimate $\tr (O \rho_\theta )$. 
Therefore, the main overhead lies in simulating real-time evolution on NQSs, for which numerous methods have been developed~\cite{Gutierrez_2022, Sinibaldi_2023, vandewalle2024many, Nys_2024_AbInitio, Gravina_2025, schmitt2025simulating, sinibaldi2026time}. 
In order to preserve quadratic error suppression, one should use a time evolution method that suffers total error $\mathcal{O}(\epsilon^2)$, especially as $|\psi_\theta\rangle$ approaches the ground state.
For example, a naive application of the time-dependent variational principle can be challenging because the ground state is a stationary point of the force vector that defines its parameter update rule~\cite{Schmitt_2020}. Later in our experiments, we use projected time-dependent VMC~\cite{Sinibaldi_2023} to better treat this issue.

\vspace{6pt}
\emph{Dynamical Averaging in the Heisenberg Picture}---
We can also simulate $\rho_\theta$ in the Heisenberg picture, where observables evolve in time. Here, Eq.~\eqref{eq:dyn_avg_eq} becomes an expectation value of a time-averaged observable: 
\begin{equation}\label{eq:dyn_avg_eq_Heis}
    \tr(O \rho_\theta) = \langle \psi_\theta | \Bigg[ \int dt \ p(t) e^{iHt} O e^{-iHt} \Bigg] |\psi_\theta \rangle . 
\end{equation}
This formulation can be realized with operator evolution methods, which can be advantageous as it avoids time-evolving a highly optimized NQS.  
Eq.~\eqref{eq:dyn_avg_eq_Heis} also reveals that dynamical averaging can only improve estimates of observables that do not commute with the Hamiltonian, since commuting observables remain invariant under time evolution. 
This is expected, as commuting observables already exhibit $\mathcal{O}(\epsilon^2)$ error (see Eq.~\eqref{eq:relative_error}). 

However, rather than implementing Eq.~\eqref{eq:dyn_avg_eq_Heis} with operator evolution, we develop an approximation that obviates time evolution altogether. Specifically, under the assumption that $|\psi_\theta\rangle$ is primarily supported on the ground state and an excited state (i.e., $|\psi_\theta \rangle \approx \sqrt{1-\epsilon^2}|E_0\rangle + \epsilon|E_1\rangle$), we derive in the Supplementary Material:
\begin{equation}\label{eq:corrected_observable}
    \tr(O\rho_\theta) \approx \langle \psi_\theta | O | \psi_\theta \rangle - \frac{ \big(\langle \psi_\theta | [H, [H,O]] | \psi_\theta \rangle \big) ^2}{ \langle \psi_\theta | [H, [H, [H, [H,O]]]] | \psi_\theta \rangle} . 
\end{equation} 
To reach this expression, we take $p(t)$ to be a zero-mean Gaussian with variance $\sigma^2$ and take the limit of large $\sigma^2$.

The second term in Eq.~\eqref{eq:corrected_observable} can be viewed as a correction to the standard observable estimate. To compute it, note that for a local observable $O$ and local Hamiltonian $H$, the nested commutators $ [H, [H,O]] $ and $[H, [H, [H, [H,O]]]]$ are both local observables, albeit with larger support than $O$. These commutators can therefore be determined analytically (e.g., with computer algebra software) and their expectation values estimated using standard NQS methods, without performing explicit real-time evolution (see Supplementary Material for details). 

We reiterate that Eq.~\eqref{eq:corrected_observable} relies on $|\psi_\theta \rangle $ being predominantly supported on two eigenstates, and its accuracy and numerical stability may deteriorate when this assumption breaks down. Heuristically, we may expect this approximation to be valid during optimization because energy minimization algorithms like stochastic reconfiguration approximate imaginary time evolution~\cite{Stokes2020quantumnatural}, which exponentially suppresses high energy eigenstates.

\vspace{6pt}
\emph{Experiments}---
Let us first exemplify dynamical averaging in the Schr\"odinger picture (Eq.~\eqref{eq:dyn_avg_eq}) on the mixed field Ising model:
\begin{equation}\label{eq:Ising_Hamiltonian}
    H_{\text{Ising}} = -J \sum_{\langle i, j\rangle } \sigma^z_i \sigma^z_{j} -h_x \sum_{i} \sigma^x_i -h_z \sum_{i} \sigma^z_i , 
\end{equation}
We set $J=1$, $h_x=1.25$, and $h_z= 0.15 $, at which point the model is non-integrable, and first consider a periodic system of $N=12$ spins to allow comparison with the exact solution. We use an NQS parametrized by a restricted Boltzmann machine with feature density $\alpha=2$ (using NetKet~\cite{netket2:2019, netket3:2022}), and employ stochastic reconfiguration to minimize energy. For dynamical averaging, we choose $p(t)$ to be a zero-mean Gaussian with variance $\sigma^2$, and perform time evolution using projected time-dependent VMC~\cite{Sinibaldi_2023} with time step size $\Delta t = 10^{-2}$. 

\begin{figure}[!t]
  \includegraphics[width=0.995\columnwidth]{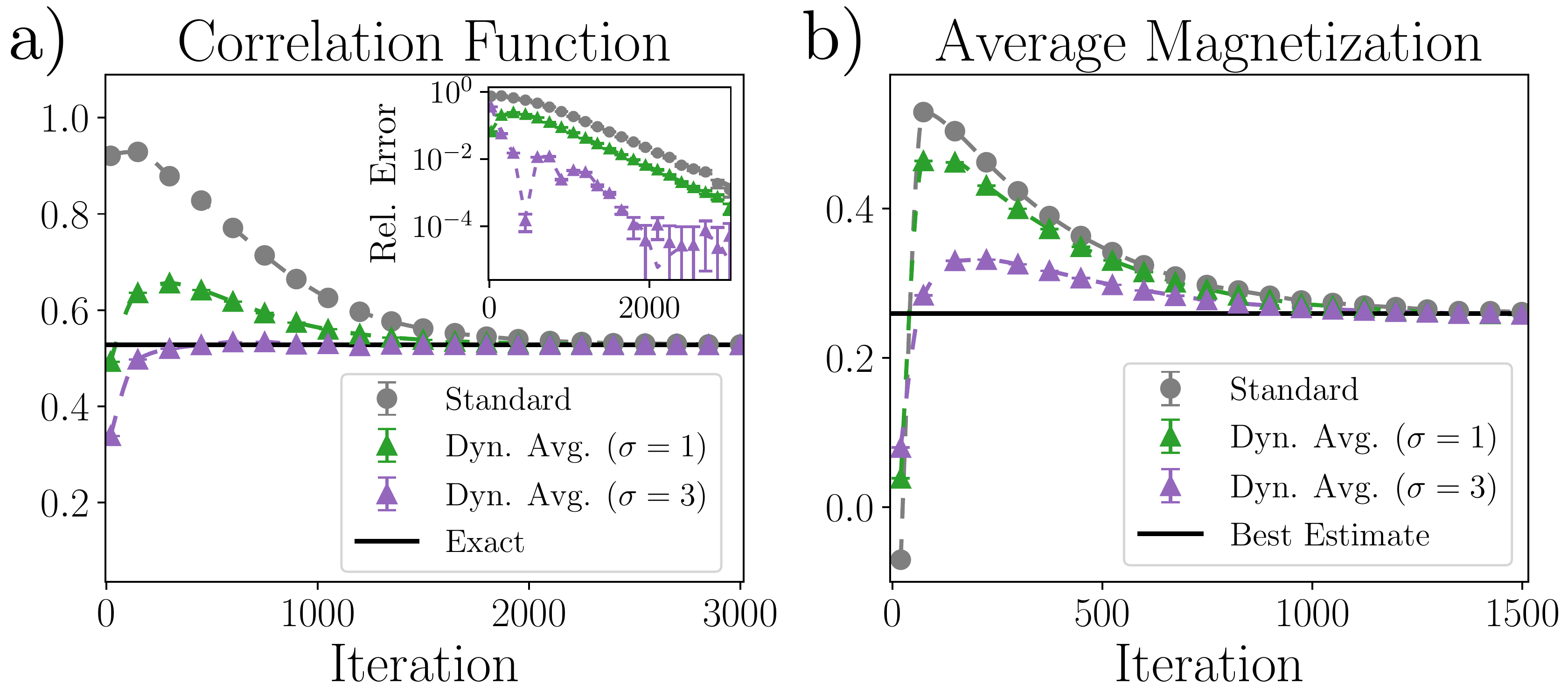}
  \caption{Comparison of standard VMC and dynamical averaging in the Schr\"odinger picture. a) The correlation function $\langle \sigma^x_{N/2} \sigma_{N/2+1}^x \rangle$ in the 1D mixed field Ising model on $N=12$ spins, including relative errors in the inset.
  b) The average magnetization $\overline{\langle \sigma^z_i\rangle}$ in the 2D mixed field Ising model on $6\times 6$ spins. 
  }
  \label{fig:experiments1}
\end{figure}

In Fig.~\ref{fig:experiments1}a, we compare the correlation function $\langle\sigma^x_{N/2} \sigma^x_{N/2+1} \rangle$ during energy minimization, evaluated with both standard VMC and dynamical averaging. Dynamical averaging consistently achieves smaller errors, with the improvement becoming more pronounced as $\sigma$ increases, in agreement with the discussion following Theorem~\ref{thm:main_theorem}. At $\sigma=1$, the relative error is reduced but still decays as $\mathcal{O}(\epsilon)$ like in standard VMC, while at $\sigma=3$ it decays as $\mathcal{O}(\epsilon^2)$ before saturating around $10^{-4}$, likely due to inaccuracies in time evolution. The dashed lines in the inset depict the exact error.

We additionally consider a regime beyond exact diagonalization, studying the same model on a $6\times 6$ square lattice with periodic boundary conditions and couplings $J=1$, $h_x = 3.3$, and $h_z = 0.05$. We follow the same procedure as above, but now using an NQS parameterized by a Vision Transformer~\cite{viteritti2023transformer, viteritti2025transformer}. We focus on the average magnetization $ \overline{\langle \sigma^z_i \rangle} = \frac{1}{N} \sum_{i=1}^{N} \langle \sigma^z_i \rangle $. Fig.~\ref{fig:experiments1}b again shows that dynamical averaging converges more quickly to the best estimate of this observable (obtained at the final iteration), with performance improving as $\sigma$ increases.

\begin{figure}[!t]
  \includegraphics[width=0.995\columnwidth]{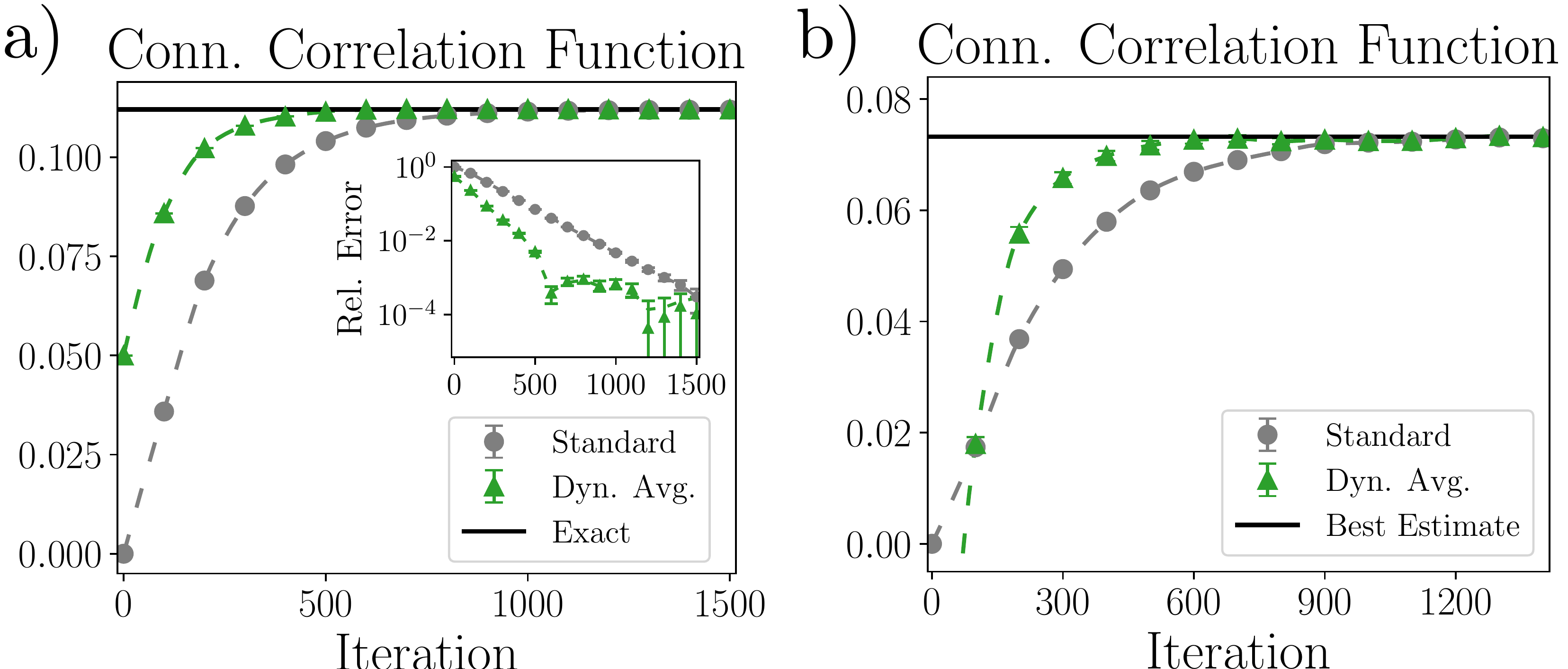}
  \caption{Comparison of standard VMC and (approximate) dynamical averaging in the Heisenberg picture.
  a) The connected correlation function $ \langle \sigma^x_{N/2} \sigma_{N/2+1}^x \rangle - \langle \sigma^x_{N/2} \rangle \langle \sigma_{N/2+1}^x \rangle$ in the 1D XXZ model on $N=12$ spins, and its relative errors.
  b) The same connected correlation function on $N=40$ spins.
  }
  \label{fig:experiments2}
\end{figure}

Next, we study our approximation to dynamical averaging in the Heisenberg picture (Eq.~\eqref{eq:corrected_observable}) applied to the 1D XXZ model with transverse magnetic field:
\begin{equation}\label{eq:XXZ_Hamiltonian}
\begin{aligned}
    H_{\text{XXZ}} =& -J\sum_{\langle i,j \rangle} \big(\sigma_i^x \sigma_{j}^x + \sigma_i^y \sigma_{j}^y + \Delta \sigma_i^z \sigma_{j}^z \big)  - h_x \sum_i \sigma_i^x .  
\end{aligned}
\end{equation}
We set $J = 1$, $\Delta = 0.25$, and $h_x = 0.25$, for which the model is non-integrable, and first consider a periodic system of $N=12$ spins to benchmark against exact diagonalization. We use an NQS parameterized by a restricted Boltzmann machine with feature density $\alpha=2$ and the SPRING optimizer~\cite{Goldshlager_2024} to minimize energy.
Fig.~\ref{fig:experiments2}a compares the connected correlation function $\langle \sigma^x_{N/2} \sigma_{N/2+1}^x \rangle - \langle \sigma^x_{N/2} \rangle \langle \sigma_{N/2+1}^x \rangle$ evaluated with standard VMC and dynamical averaging. As evidenced in this plot, dynamical averaging suppresses relative errors by about an order of magnitude during optimization and accelerates convergence to the exact value. The relative error ultimately saturates around $10^{-4}$, limited by the accuracy of the approximation used to derive Eq.~\eqref{eq:corrected_observable}. 

Lastly, we study this system beyond exact diagonalization by considering the same model on $N=40$ spins, now with parameters $\Delta = 0.5$ and $h_x = 0.2$. Using the same approach as above, Fig.~\ref{fig:experiments2}b shows the connected correlation function during optimization, again demonstrating that dynamical averaging achieves faster convergence to the best estimate of this quantity.

Overall, these results show that dynamical averaging reduces errors and accelerates convergence, even in large systems. Nevertheless, the method has its limitations.
Its realization in the Schr\"odinger picture requires evolving an NQS in real-time, which can be computationally costly in challenging systems. In contrast, our approach in the Heisenberg picture is much cheaper to evaluate, but relies on the assumption that $|\psi_\theta \rangle$ is predominantly supported on a few low energy eigenstates. When this assumption is violated, this approach can experience additional errors and numerical instabilities. 
An interesting direction for future work is to refine dynamical averaging in the Heisenberg picture using operator evolution methods, like Pauli propagation~\cite{Rall_Simulation_2019, rudolph2025pauli}.

\vspace{6pt}
\emph{Conclusions}---
We have used insights from quantum information to develop a new NQS algorithm that improves ground state approximations. At a conceptual level, dynamical averaging exploits a fundamental feature of quantum mechanics: a quantum state is really quadratic in the state vector, i.e. $\rho = |\psi\rangle \langle \psi |$, whereas typical NQS methods optimize only the state vector $|\psi_\theta\rangle$. By averaging over many such state vectors, we promote the variational object to a mixed state that can better approximate the ground state. This adds to the growing body of research that uses randomization to improve quantum state preparation and estimation~\cite{harrow2025randomized, wang2025faster, arulandu2026trading, zlokapa2026optimal, watson2025randomly, angrisani2026}.

Our work suggests several natural extensions. 
For one, dynamical averaging could be applied to excited state NQSs~\cite{Entwistle_2023, Pfau_2024}, or extended to other variational methods like tensor networks~\cite{Sandvik_2007, Schuch_2008, Ferris_2012, Liu_2021, Liu_2024, Liu_2025} and Jastrow ans\"atze~\cite{Foulkes_2001, Sorella_2005, Motta_2023}. 
Beyond classical simulation, the ideas developed here may also find use in related settings, such as in variational quantum algorithms~\cite{peruzzo2014variational, 2021Nat_Variational, Tilly_2022} or machine learning problems that optimize an energy-like quantity. 

Moreover, dynamical averaging can be generalized beyond real-time dynamics. In general, any operation that acts differently on the eigenstates of $H$ can, when appropriately randomized, suppress contributions from undesired states. For instance, symmetry operators, such as translations or rotations, can be randomly applied to selectively target states in certain symmetry sectors~\cite{greene2021error}. 

From a quantum information perspective, randomizing over symmetry operators to isolate a desired symmetry sector parallels the stabilizer formalism of quantum error correction~\cite{gottesman1997stabilizer, Terhal_2015, greene2021error}. However, unlike error correction, our procedure constitutes only a form of error mitigation, which reduces errors but cannot suppress them arbitrarily.
Nonetheless, this connection suggests that NQSs, and classical simulation methods in general, could benefit considerably from a synthesis with quantum information, which provides both the conceptual and technical tools to design better simulation algorithms.

\vspace{6pt}
\emph{Code Availability}---
Code to run the experiments conducted in
this paper is available on GitHub at \url{https://github.com/rr637/DynamicalAveragingNQS}.

\vspace{6pt}
\emph{Acknowledgments}---
JMM acknowledges insightful conversations with Alev Orfi, Luciano Viteritti, Giuseppe Carleo, Dries Sels, Norm Tubman, Samuel Buckley-Bonanno, Noah Eckstein, and Sam Garratt. 

This research was supported by PNNL’s Quantum Algorithms and Architecture for Domain Science (QuAADS) Laboratory Directed Research and Development Initiative. 
This research was supported by the Physical and Computational Sciences Directorate Seed, under the Laboratory Directed Research and Development Program at PNNL. 
This material is based upon work supported by the U.S. Department of Energy, Office of Science, National Quantum Information Science Research Centers, Quantum Science Center (QSC). 
PNNL is a multi-program national laboratory operated for the U.S. Department of Energy (DOE) by Battelle Memorial Institute under Contract No. DE-AC05-76RL01830.

\bibliography{Refs}

\appendix
\clearpage 

\onecolumngrid
\begin{center}
	\noindent\textbf{Supplementary Material}
	\bigskip
	\noindent\textbf{\large{}}
\end{center}
\twocolumngrid

\section{Proof of Theorem~\ref{thm:main_theorem}}\label{app:proof}
\begin{proof}
     For simplicity, we will take the ground state energy to be zero, $E_0 = 0$, and make appropriate replacements at the end of the proof to account for a non-zero ground state energy. Consequently, the excitation energies are $\Delta_n = E_n - E_0 = E_n$. This effectively shifts the energy by a constant, which only changes physically-irrelevant global phases.
     
     To begin, let us expand $|\psi_\theta\rangle$ in the eigenbasis of $H$ as $|\psi_\theta\rangle = \sum_{n} a_n |E_n\rangle$, where we take $a_0 \in \mathbb{R}$ and $a_0 > 0$ without loss of generality. Then, for pure states $|\psi_\theta\rangle $ and $|E_0\rangle$, their trace distance evaluates to $d_{\text{tr}}(|\psi_\theta\rangle \langle \psi_\theta| , |E_0\rangle \langle E_0 | ) = \sqrt{1-|\langle E_0 |\psi_\theta\rangle |^2 } \leq \epsilon$~\cite{wilde2013quantum}, which implies that $a_0 = \langle E_0 | \psi_\theta\rangle \geq \sqrt{1-\epsilon^2} = 1- \mathcal{O}(\epsilon^2)$. 
    
    We can then prove the theorem by direct computation. Let us denote the time-evolved state as $|\psi_\theta(t)\rangle = e^{-iHt}|\psi_\theta\rangle$, and the density matrix of interest as $\rho_\theta = \mathbb{E}_{p(t)} \big[ |\psi_\theta(t) \rangle  \langle \psi_\theta(t) | \big]$. Let us also define $|\delta(t)\rangle := |\psi_\theta(t)\rangle - |E_0\rangle $. We can express the 1-norm that appears in $d_{\text{tr}}(\rho_\theta, |E_0\rangle \langle E_0|) $ as 
    \begin{equation}\label{eq:1_norm}
    \begin{aligned}
         &\big\| \rho_\theta - |E_0\rangle \langle E_0| \big\|_1 \\
        & \ = \Big\| \mathbb{E} \big[ |\psi_\theta(t)\rangle \langle \psi_\theta(t) | - |E_0 \rangle \langle E_0 | \big] \Big\|_1 \\
        &\ = \Big\| \mathbb{E} \big[ |\delta(t) \rangle \langle \delta (t) | + |\delta(t)\rangle \langle E_0| + |E_0\rangle \langle \delta(t) |  \big] \Big\|_1 \\
        & \ \leq \Big\| \mathbb{E} \big[ |\delta(t) \rangle \langle \delta(t)| \big] \Big\|_1 + \Big\| \mathbb{E} \big[ | \delta(t)\rangle \big] \langle E_0 |  \Big\|_1 \\
        & \qquad \qquad \qquad \qquad \quad \ + \Big\|  | E_0\rangle \cdot \mathbb{E} \big[ \langle \delta(t) | \big] \Big\|_1 
    \end{aligned}
    \end{equation}
    where we have invoked the triangle inequality in obtaining the last line.

    The first term is upper bounded as $\big\| \mathbb{E} \big[ |\delta(t) \rangle \langle \delta(t)| \big] \big\|_1 \leq \mathbb{E} \big[ \| |\delta(t) \rangle \langle \delta(t)| \|_1 \big] $. To evaluate this, recall that the 1-norm is the sum of the singular values. Noting that $|\delta(t) \rangle \langle \delta(t)|$ is itself an un-normalized singular value decomposition, its 1-norm is the squared 2-norm of $|\delta(t)\rangle$:
    \begin{equation}
        \big\| |\delta(t) \rangle \langle \delta(t)| \big\|_1 = 
        \langle \delta(t) |\delta(t) \rangle = 
        2-2a_0 \leq \mathcal{O}(\epsilon^2) . 
    \end{equation}
    Therefore, we have that $\big\| \mathbb{E} \big[ |\delta(t) \rangle \langle \delta(t)| \big] \big\|_1 \leq \mathcal{O}(\epsilon^2)$.  
    
    The next term in Eq.~\eqref{eq:1_norm} similarly evaluates to
    \begin{equation}
    \begin{aligned}
        \Big\| \mathbb{E} \big[ | \delta(t)\rangle \big] \langle E_0 |  \Big\|_1 &= \big\| \mathbb{E} [\delta(t) \rangle ] \big\|_2 \\
        &= \Big\| |E_0\rangle - \int dt p(t) e^{-iHt}|\psi_\theta\rangle \Big\|_2.
    \end{aligned}
    \end{equation}
    Inserting $|\psi_\theta\rangle = \sum_n a_n |E_n\rangle$, and using the orthogonality of the eigenstates, this norm evaluates to 
    \begin{equation}
    \begin{aligned}
        & \Big[ 1-2a_0 +\sum_n |a_n|^2 \int dt dt' p(t) p(t') e^{-iE_n (t-t')} \Big]^{1/2} \\
        &\ = \Big[ 1-2a_0 + a_0^2 + \sum_{n\geq 1} |a_n|^2 |\varphi(E_n)|^2 \Big]^{1/2} \\
        & \ \leq \Big[ (1-a_0)^2 + \Big(\sum_{n\geq 1} |a_n|^2 \Big)\cdot \max_{n\geq 1} |\varphi(E_n)|^2 \Big]^{1/2} \\
        & \ \leq \Big[ \mathcal{O}(\epsilon^4) + \mathcal{O}\big( \epsilon^2 \cdot \max_{ n\geq 1} |\varphi(E_n)|^2 \big) \Big]^{1/2} \\
        & \ = \mathcal{O}(\epsilon^2) + \mathcal{O}\big( \epsilon\cdot \max_{ n\geq 1} |\varphi(E_n)| \big)
    \end{aligned}
    \end{equation}
    where $\varphi(E_n) = \int dt \ p(t) e^{iE_n t}$ is the characteristic function of $p(t)$, and we have used that $\sum_{n\geq 1} |a_n|^2 = 1-a_0^2 \leq \mathcal{O}(\epsilon^2)$. Furthermore, the third term in Eq.~\eqref{eq:1_norm} is analogously upper bounded.

    Collecting all the above terms, we have:
    \begin{equation}
        \big\| \rho_\theta- |E_0\rangle \langle E_0| \big\|_1 \leq \mathcal{O}(\epsilon^2) + \mathcal{O} \big( \epsilon\cdot \max_{n\geq 1} |\varphi(E_n)| \big) . 
    \end{equation}
    Then, reintroducing a non-zero ground state energy $E_0$ amounts to shifting the energies $E_n$ to excitation energies $E_n - E_0 = \Delta_n$. This yields the stated bound:
    \begin{equation}
    \begin{aligned}
        d_{\text{tr}}\big( \rho_\theta, \ |E_0\rangle \langle E_0 | \big) &= \frac{1}{2} \big\| \rho_\theta- |E_0\rangle \langle E_0| \big\|_1 \\
        &\leq \mathcal{O}(\epsilon^2) + \mathcal{O} \big( \epsilon\cdot \max_{n\geq 1} |\varphi(\Delta_n)| \big) . 
    \end{aligned}
    \end{equation}
    For completeness, we note that the same result can alternatively be derived using the concept of \textit{randomized compiling}~\cite{hastings2016turning, Campbell_2017}, as is often considered in quantum computing contexts. 
    
\end{proof}

\section{High-Accuracy NQSs in the Heisenberg Picture}

As we discussed in the main text, a straightforward way to realize dynamical averaging is by time-evolving the state vector as $|\psi_\theta(t)\rangle = e^{-iHt} |\psi_\theta\rangle $. This corresponds to the \textit{Schr\"odinger picture} of quantum mechanics, where states evolve and observables remain fixed. Alternatively, dynamical averaging can also be realized in the \textit{Heisenberg picture}, where states are static and observables evolve in time. This approach can be more favorable in classical simulation where directly time-evolving wave functions can be costly, in contrast to simulation on quantum computers where Schr\"odinger-picture time evolution is more efficient.

To develop dynamical averaging in the Heisenberg picture, consider estimating an observable in the mixed state from Eq.~\eqref{eq:rho} in the main text: $\rho_\theta = \int dt p(t) e^{-iHt} |\psi_\theta\rangle \langle \psi_\theta | e^{iHt}$. We can express an observable as
\begin{equation}\label{eq:dyn_avg_obs_appendix}
\begin{aligned}
    \tr(O\rho_\theta) & = \int dt \ p(t) \langle \psi_\theta | e^{iHt} O e^{-iHt} |\psi_\theta \rangle \\
    &= \langle \psi_\theta | \bigg[ \int dt \ p(t)  e^{iHt} O e^{-iHt} \bigg] |\psi_\theta \rangle. 
\end{aligned}
\end{equation}
Using the nested commutator expansion of a time-evolved operator, we write 
\begin{equation}
    e^{iHt} O e^{-iHt} = e^{it [H, \cdot]} (O),
\end{equation}
where $e^{it [H, \cdot]}$ is a superoperator defined by the Taylor expansion: 
\begin{equation}\label{eq:time_evolved_obs}
\begin{aligned}
    e^{it [H, \cdot]} &= \sum_{n=0}^\infty \frac{(it)^n [H, \cdot]^n}{n!} \\
    &= I + it [H,\cdot] - \frac{t^2}{2!} [H,[H,\cdot]] + \ldots  
\end{aligned}
\end{equation}
Here the power notation denotes a nested commutator:
\begin{equation}\label{eq:nested_comm_notation}
\begin{aligned}
     &[H, \cdot]^n = \underbrace{[H,[H, ... [H,\cdot]]]}_{n\text{ times}} . 
\end{aligned}
\end{equation}
Substituting this into the integral in Eq.~\eqref{eq:dyn_avg_obs_appendix}, we obtain
\begin{equation}\label{eq:proxy_obs_integral_eq}
\begin{aligned}
    \int dt\  p(t) e^{itH}O e^{-itH} &= 
    \int dt \ p(t) e^{it [H, \cdot]}(O) \\
    &= \varphi([H,\cdot])(O) , 
\end{aligned}
\end{equation}
where $\varphi(x) = \int dt p(t) e^{ixt}$ is the characteristic function of $p(t)$. The superoperator $ \varphi([H,\cdot])$ is defined by the Taylor series of $\varphi(x)$ (assuming this series exists and converges), analogous to Eq.~\eqref{eq:time_evolved_obs}. That is, if $\varphi(x) = \sum_{j=0}^\infty b_j x^j $, then $\varphi([H,\cdot]) = \sum_{j=0}^\infty b_j [H, \cdot ]^j $.

Altogether, we can express an observable as
\begin{equation}\label{eq:dyn_avg_proxy_obs}
    \tr(O\rho_\theta) = \langle \psi_\theta | \varphi([H,\cdot ])(O) |\psi_\theta \rangle. 
\end{equation} 
Eq.~\eqref{eq:dyn_avg_proxy_obs} contrasts conventional estimation of a ground state observable, in which one directly evaluates $\langle \psi_\theta | O | \psi_\theta \rangle$. Rather, Eq.~\eqref{eq:dyn_avg_proxy_obs} replaces the observable with the operator $\varphi([H,\cdot ])(O)$, which acts as a proxy for the observable $O$. As this formulation is equivalent to dynamical averaging in the Schr\"odinger picture, it inherits the same results of the main text, including the quadratic suppression of error for general observables.

The operator $\varphi([H,\cdot ])(O)$ has many favorable properties. By Eq.~\eqref{eq:proxy_obs_integral_eq}, it is clearly Hermitian because it is an average of Hermitian operators. In addition, because $[H+k, O] = [H,O]$ for any scalar constant $k$, it is also invariant under shifting $H$ by a constant. Furthermore, $\varphi(0) = 1$ for any distribution $p(t)$, so the first term in the expansion of $\varphi([H,\cdot])(O)$ is $O$:
\begin{equation}
    \varphi([H,\cdot])(O) =  O + \sum_{n=1}^\infty b_j [H, \cdot]^j(O), 
\end{equation}
where $b_j$ are the coefficients of this expansion. The higher order terms can be interpreted as corrections to $O$ that improve the approximation to the ground state observable $\langle E_0 | O |E_0\rangle$.

To better understand the action of $\varphi([H,\cdot])(O)$, let us consider how it acts on the eigenstates of the Hamiltonian. In particular, let us consider the matrix element 
\begin{equation}
    \langle E_{n'}| \varphi([H,\cdot])(O) |E_{n}\rangle . 
\end{equation}
Using the shift invariance of $\varphi([H,\cdot])(O)$ noted above, let us write $\varphi([H,\cdot])(O) = \varphi([H-E_{n'},\cdot])(O)$. Writing the characteristic function as $\varphi(x) = \sum_{j=0}^\infty b_j x^j$, we have
\begin{equation}
\begin{aligned}
    &\langle E_{n'} | \varphi([H, \cdot])(O) | E_n \rangle \\
    &=\langle E_{n'} | \varphi([H-E_{n'},\cdot])(O) | E_n \rangle \\ 
    &=\sum_j b_j \langle E_{n'} |  [H-E_{n'},\cdot]^j (O) | E_n \rangle . 
\end{aligned}
\end{equation}
To evaluate this, note that in the expansion of the nested commutator $[H-E_{n'},\cdot]^j (O)$, all terms with a factor of $H - E_{n'}$ to the left of $O$ (e.g, $(H-E_{n'})^2 O (H-E_{n'})^{j-2}$) annihilate the state $\langle E_{n'} |$ and therefore vanish. The only term that remains is $O(H-E_{n'})^j$. Therefore, we have
\begin{equation}
\begin{aligned}
    &\langle E_{n'} | \varphi([H, \cdot])(O) | E_n \rangle  \\
    &=\sum_n b_n \langle E_{n'} | O (H-E_{n'})^j | E_n \rangle \\
    &=\sum_j b_j (E_n - E_{n'})^j \langle E_{n'} | O | E_n \rangle \\
    &= \varphi(E_n - E_{n'}) \langle E_{n'} | O | E_n \rangle . 
\end{aligned}
\end{equation}

We can use this to express the expectation value of the operator $\varphi([H, \cdot])(O)$ on a state $|\psi_\theta\rangle = \sum_n a_n |E_n\rangle$ as
\begin{equation}\label{eq:proxy_obs_eval}
\begin{aligned}
    &\langle \psi_\theta | \varphi([H, \cdot])(O) |\psi_\theta \rangle \\
    &=\sum_{n,n'} a_n a_{n'}^* \langle E_{n'} | \varphi([H, \cdot])(O) |E_n \rangle \\
    &= \sum_n |a_n|^2 \langle E_n | O | E_n \rangle \\
    & \qquad \quad + \sum_{n,n'; \ n\neq n'} a_n a_{n'}^* \varphi(E_n - E_{n'} ) \langle E_{n'} | O | E_n \rangle , 
\end{aligned}
\end{equation}
where we have used $\varphi(0) = 1$. Compared to the standard expression for $\langle \psi_\theta | O | \psi_\theta \rangle$, Eq.~\eqref{eq:proxy_obs_eval} leaves the diagonal contributions $\langle E_n | O | E_n \rangle $ unchanged, but modifies the off-diagonal terms $\langle E_{n'} | O | E_n \rangle $ by a factor of $\varphi(E_n - E_{n'} )$.

As we noted in the main text Eq.~\eqref{eq:relative_error}, the dominant error in ground state observables arises from off-diagonal contributions, which are precisely the terms that dynamical averaging seeks to suppress. By Theorem~\ref{thm:main_theorem}, we can achieve this suppression by choosing a probability distribution $p(t)$ whose characteristic function $\varphi(E_n - E_{n'})$ is sufficiently small for $n \neq n'$. Eq.~\eqref{eq:proxy_obs_eval} makes this connection explicitly clear: $\varphi(E_n - E_{n'} )$ multiplies the off-diagonal contributions $\langle E_{n'} | O | E_n \rangle $ and thus suppresses their contribution to the error. For example, in the main text we assumed the condition $d_{\text{tr}} \big( |\psi_\theta\rangle \langle \psi_\theta | , \ |E_0\rangle \langle E_0| \big) \leq \epsilon$, which implies that $|a_n a_{n'}| \leq \mathcal{O}(\epsilon)$ for all pairs $n,n'$. Therefore, it suffices to choose $\varphi(E_n - E_{n'}) \leq \mathcal{O}(\epsilon)$ to guarantee $\mathcal{O}(\epsilon^2)$ relative error in estimating $\langle E_0 | O | E_0 \rangle$.

Beyond this illuminating perspective provided by the Heisenberg picture, it also offers computational methods for implementing dynamical averaging. By Eq.~\eqref{eq:dyn_avg_obs_appendix}, the expectation value $\langle \psi_\theta | \varphi([H,\cdot]) (O) | \psi_\theta\rangle$ is an average over time-evolved observables $ e^{iHt} O e^{-iHt}$, which can be evaluated using operator-evolution methods like Pauli propagation methods~\cite{Rall_Simulation_2019, rudolph2025pauli}. Moreover, in a later appendix section, we use the Heisenberg picture approach to derive an approximation to $\langle \psi_\theta | \varphi([H,\cdot]) (O) | \psi_\theta\rangle$ as a truncated expansion of nested commutators. This approximation is particularly useful because it obviates the need to perform explicit time evolution.

\section{Analogy Between Dynamical Averaging and Imaginary Time Evolution}
We note that the above analysis in the Heisenberg picture also indicates that dynamical averaging is similar to, but distinct from, imaginary time evolution. Consider taking the distribution $p(t)$ to be a zero-mean Gaussian of width $\sigma$ such that its characteristic function is $\varphi(x) = e^{-\sigma^2 x^2/2}$, and thus $\varphi([H,\cdot])(O) = e^{-\frac{\sigma^2}{2} [H,[H,\cdot]]}(O)$.\footnote{Curiously, this operator is also the solution to the differential equation $\frac{d}{ds} \varphi(s) = -\sigma^2 [H,[H,\varphi(s)]]$ evaluated at $s=1$, which arises in studies of operator evolution under Brownian motion and has uses in quantum simulation of thermal states~\cite{buckleybonanno2026quantumsimulationgaugetheories}. } Then, computing the expectation value $\langle \psi_\theta |e^{-\frac{\sigma^2}{2} [H,[H,\cdot]]}(O) | \psi_\theta\rangle $ appears analogous to applying imaginary time evolution to $|\psi_\theta\rangle$ for a time $\sigma^2/2$.

However, while both imaginary time evolution and dynamical averaging produce a finer approximation to ground state observables, they do so through different mechanisms. Imaginary time evolution suppresses all excited state amplitudes in $| \psi_\theta \rangle$, whereas dynamical averaging suppresses the off-diagonal elements of its density matrix in the energy eigenbasis. Moreover, imaginary time evolution requires renormalization of the state after its application, whereas dynamical averaging preserves the normalization of the state.

\section{High-Accuracy NQSs in the Thermodynamic Limit}

In the main text, Theorem~\ref{thm:main_theorem} specialized to the case where an NQS is close in trace distance to the ground state. While this holds in finite sized systems, this intuition can break down in the thermodynamic limit due to the orthogonality catastrophe, where the fidelity between the variational state and ground state can be small (typically decaying with system size), even as local observables converge to their ground state values. Nevertheless, there still remains a disparity between the errors in the energy and arbitrary observables.

To analyze this, let us consider a state $|\psi_\theta\rangle = \sum_n a_n |E_n\rangle$. The relative error in an observable $O$ is
\begin{equation}\label{eq:generic_rel_err}
\begin{aligned}
    &\Bigg| \frac{ \langle \psi_\theta | O | \psi_\theta \rangle - \langle E_0 | O | E_0 \rangle }{ \langle E_0 | O | E_0 \rangle } \Bigg| \\
    =&\Bigg| \big(|a_0|^2 - 1 \big) +  \sum_{n\geq 1} |a_n|^2 \frac{\langle E_{n}| O | E_n\rangle}{\langle E_0 | O | E_0\rangle} \\
    & \qquad \qquad \ \ \  + \sum_{n \neq n'} a_n a_{n'}^* \frac{\langle E_{n'}| O | E_n\rangle}{\langle E_0 | O | E_0\rangle} \Bigg| \\
    =& \Bigg| \sum_{n\geq 1} |a_n|^2 \bigg( \frac{\langle E_{n}| O | E_n\rangle}{\langle E_0 | O | E_0\rangle} -1 \bigg) + \sum_{n \neq n'} a_n a_{n'}^* \frac{\langle E_{n'}| O | E_n\rangle}{\langle E_0 | O | E_0\rangle} \Bigg| , 
\end{aligned}
\end{equation}
where we have used $|a_0|^2 = 1-\sum_{n\geq 1} |a_n|^2$. Thus, for an arbitrary observable where $\langle E_{n'}| O | E_n\rangle \neq 0$ in general, its error receives contributions from both the diagonal and off-diagonal elements in the energy eigenbasis. In contrast, for the energy, the off-diagonal contributions vanish ($\langle E_{n'} |H|E_n\rangle = 0 $ for $n\neq n'$), so its error depends only on the diagonal contributions. Qualitatively, this explains why even in large systems, the energy can converge faster than a generic observable, although the fidelity with the ground state may remain small.

We can use this reasoning to show that dynamical averaging still suppresses errors in these large systems, where assumptions on the trace distance with the ground state may not be appropriate. Eqs.~\eqref{eq:dyn_avg_proxy_obs} and~\eqref{eq:proxy_obs_eval} indicate that an observable computed with dynamical averaging suffers relative error
\begin{equation}
\begin{aligned}
    & \Bigg| \sum_{n\geq 1} |a_n|^2 \bigg( \frac{\langle E_{n}| O | E_n\rangle}{\langle E_0 | O | E_0\rangle} -1 \bigg) \\
    & \qquad \quad + \sum_{n \neq n'} a_n a_{n'}^* \varphi(E_n-E_{n'}) \frac{\langle E_{n'}| O | E_n\rangle}{\langle E_0 | O | E_0\rangle} \Bigg| . 
\end{aligned}
\end{equation}
Evidently, the off-diagonal terms are suppressed by a factor of $\varphi(E_{n'} - E_{n})$. This can be seen as analogous to a dephasing channel (or phase damping channel)~\cite{nielsen2000quantum}, which similarly suppresses off-diagonal elements of the density matrix. For an appropriately chosen probability distribution $p(t)$ whose characteristic function strongly suppresses these terms, only the diagonal terms contribute significantly to the error, yielding the same error scaling as the energy.

Therefore, in large systems the advantage of dynamical averaging is more naturally understood not as a quadratic suppression of the trace distance, but as a suppression of the off-diagonal contributions to the error in an observable. This provides a qualitative improvement in error scaling and enables one to estimate observables with the same accuracy as the energy. A precise quantitative improvement could be established under additional assumptions on the size of the diagonal and off-diagonal contributions.

\section{Deriving an Approximation to Dynamical Averaging in the Heisenberg Picture}

In this appendix, we derive an approximation to dynamical averaging in the Heisenberg picture that obviates the need to perform explicit time evolution. To begin, let us again take the distribution $p(t)$ to be a Gaussian of width $\sigma$ such that its characteristic function is $\varphi(x) = e^{-\sigma^2 x^2/2}$, and so $\varphi([H,\cdot])(O) = e^{-\frac{\sigma^2}{2} [H,[H,\cdot]]}(O)$. Also, let the variational state be $|\psi_\theta\rangle = \sum_n a_n |E_n\rangle$. Then, according to the discussion following Eq.~\eqref{eq:proxy_obs_integral_eq}, we have that 
\begin{equation}\label{eq:Proxy_O_series}
\begin{aligned}
    \tr( O \rho_\theta ) &= \langle \psi_\theta | \varphi([H,\cdot])(O) |\psi_\theta \rangle \\
    &= \sum_{n=0}^\infty \Big( \frac{-\sigma^2}{2} \Big)^n \frac{1}{n!} \langle \psi_\theta | [H,\cdot]^{2n}(O) |\psi_\theta \rangle \\
    &= \sum_{n=0}^\infty \Big( \frac{-\sigma^2}{2} \Big)^n \frac{1}{n!} M_{2n}(O) , 
\end{aligned}
\end{equation}
where we have defined the expectation value of the $k$th-order nested commutator as
\begin{equation}
    M_{k}(O) := \langle \psi_\theta | [H,\cdot]^{k}(O) |\psi_\theta \rangle .  
\end{equation}
Next, note that by Eq.~\eqref{eq:proxy_obs_eval}, we can equivalently express this quantity as 
\begin{equation}\label{eq:Proxy_O_eigenstates}
\begin{aligned}
    &\langle \psi_\theta | \varphi([H,\cdot])(O) |\psi_\theta\rangle \\
    &= \sum_j |a_n|^2 \langle E_n | O |E_n\rangle \\
    &\qquad \qquad + \sum_{n \neq n'} a_n a_{n'}^* e^{-\frac{\sigma^2}{2} (E_n - E_{n'})^2} \langle E_n |O|E_{n'} \rangle \\
    &= D(O) + \sum_{n \neq n'} a_n a_{n'}^* e^{-\frac{\sigma^2}{2} (E_n - E_{n'})^2} \langle E_n |O|E_{n'} \rangle , 
\end{aligned}
\end{equation}
where we have defined $D(O)$ as the sum of the diagonal contributions 
\begin{equation}
    D(O) := \sum_n |a_n|^2 \langle E_n | O |E_n\rangle . 
\end{equation}
As $\sigma \rightarrow \infty$, $\langle \psi_\theta | \varphi([H,\cdot])(O) |\psi_\theta \rangle$ converges to $D(O)$. Thus $D(O)$ can be viewed as the best observable estimate obtained from dynamical averaging, corresponding to the complete suppression of all off-diagonal contributions. This provides the best approximation to the ground state observable $\langle E_0 | O | E_0 \rangle$ without modifying the state $|\psi_\theta \rangle$.

While exactly evaluating $\langle \psi_\theta | \varphi([H,\cdot] ) (O) | \psi_\theta \rangle$ generally requires real-time evolution, we can develop an approximation to it that avoids explicit time evolution as follows. First, we make the assumption that $| \psi_\theta \rangle$ is close to the ground state, and that its deviation from the ground state is mostly concentrated on the first excited state (or really, any single excited state):
\begin{equation}\label{eq:approx_GS}
    |\psi_\theta \rangle \approx a_0 |E_0\rangle + a_1 |E_1\rangle , 
\end{equation}
where $|a_1| \ll |a_0| $. Heuristically, we expect this condition to be met as $|\psi_\theta\rangle$ converges to the ground state under energy minimization. Under this approximation, the equality of Eqs.~\eqref{eq:Proxy_O_series} and~\eqref{eq:Proxy_O_eigenstates} indicates that 
\begin{equation}
\begin{aligned}
    &\langle \psi_\theta | \varphi([H,\cdot])(O) |\psi_\theta \rangle \\
    &= \sum_{n=0}^\infty \Big( \frac{-\sigma^2}{2} \Big)^n \frac{1}{n!} M_{2n}(O) \\
    &\approx  D(O) + e^{-\frac{\sigma^2}{2} \Delta^2} \cdot 2\text{Re} \Big( a_0 a_{1}^* \langle E_0 |O|E_{1} \rangle \Big) \\
    &=  D(O) + e^{-\frac{\sigma^2}{2} \Delta^2} \cdot \gamma
\end{aligned}
\end{equation}
where $\Delta = E_1 - E_0$ is the spectral gap, and we have defined $\gamma := 2 \text{Re} \big( a_0 a_{1}^* \langle E_0 |O|E_{1} \rangle \big)$. By expanding $e^{-\sigma^2 x^2/2}$ in powers of $\sigma^2$, this expression translates to 
\begin{equation}
\begin{aligned}
    & M_{0}(O) - \frac{\sigma^2}{2} M_2(O) + \frac{1}{2}\Big( \frac{\sigma^2}{2} \Big)^2  M_4(O) + ... \\
    &\approx  D(O) + \gamma \Big(1 - \frac{\sigma^2}{2} \Delta^2 + \frac{1}{2}\Big(\frac{\sigma^2}{2}\Big)^2 \Delta^4 + ... \Big)  .  
\end{aligned}
\end{equation}
Matching powers of $\sigma$, we find 
\begin{equation}
    M_0(O) = D(O) +\gamma , \ \  M_2(O) =  \gamma \Delta^2, \ \ M_4(O) = \gamma \Delta^4 . 
\end{equation}
This is a system of equations that can be solved to yield 
\begin{equation}
    \Delta^2 = M_4(O)/M_2(O), \ \ \gamma = M_2(O)^2 / M_4(O) ,  
\end{equation}
from which $D(O)$ can be extrapolated as
\begin{equation}
    D(O) = M_0(O) - \gamma = M_0(O) - \frac{M_2(O)^2}{M_4(O)} . 
\end{equation}
Inserting the expression for $M_k(O)$, we find that our approximation of the ground state observable $\langle E_0| O |E_0\rangle $ is 
\begin{equation}\label{eq:corrected_observable_app}
    \langle \psi_\theta | O | \psi_\theta \rangle - \frac{ \big(\langle \psi_\theta | [H, [H,O]] | \psi_\theta \rangle \big) ^2}{ \langle \psi_\theta | [H, [H, [H, [H,O]]]] | \psi_\theta \rangle  } . 
\end{equation}

To understand this expression, we can view the term $-\frac{ (\langle \psi_\theta | [H, [H,O]] | \psi_\theta \rangle ) ^2}{ \langle \psi_\theta | [H, [H, [H, [H,O]]]] | \psi_\theta \rangle  }$ as a correction to the ordinary observable estimate $\langle \psi_\theta | O | \psi_\theta \rangle$. 
To understand this correction term, note that both $ [H, [H,O]] $ and $[H, [H, [H, [H,O]]]]$ are Hermitian (for any Hermitian $O$), which implies that their expectation values---and thus also the correction term---are manifestly real-valued. Moreover, when $O$ is a local observable and $H$ is a sum of local interactions, so too will $[H, [H,O]] $ and $[H, [H, [H, [H,O]]]] $ be local Hermitian observables, albeit with larger support than $O$. In practice, we can evaluate these nested commutators analytically given a Hamiltonian $H$ and local observable $O$ (for instance, using computer algebra software). This means that we can evaluate their expectation values $\langle \psi_\theta | [H, [H,O]] | \psi_\theta \rangle $ and $\langle \psi_\theta | [H, [H, [H, [H,O]]]] | \psi_\theta \rangle$ on an NQS using standard VMC methods, without having to explicitly perform real-time evolution.

The correction term also displays other favorable properties. First, for any observable that commutes with the Hamiltonian, the corresponding correction term vanishes because $[H,O] = 0$. This is consistent with the fact that dynamical averaging only improves the accuracy of observables that do not commute with the Hamiltonian, as an observable that commutes with the Hamiltonian will already display $\mathcal{O}(\epsilon^2)$ error scaling as $|\psi_\theta\rangle$ converges to the ground state (see Eq.~\eqref{eq:relative_error} and the surrounding discussion). Second, as $|\psi_\theta\rangle$ converges to the ground state, the correction vanishes. This is because $H$ acts as a constant on the ground state ($H|E_0\rangle = E_0 |E_0\rangle$), and thus the expectation value of a commutator of $H$ acting on the ground state is necessarily zero: $\langle E_0 | [H, [H,O]] | E_0  \rangle = 0 = \langle E_0 | [H, [H, [H, [H,O]]]] | E_0 \rangle$. This is consistent with the fact that as $|\psi_\theta \rangle$ converges to the ground state, the standard expectation value $\langle \psi_\theta | O | \psi_\theta \rangle$ converges to $\langle E_0 | O | E_0 \rangle $, and so the correction should vanish. Lastly, we note that in both of these cases, the expectation values in the numerator and denominator of the correction term both vanish, which might appear to introduce a potential singularity. However, because the expectation value in the numerator is squared (see Eq.~\eqref{eq:corrected_observable_app}), the numerator vanishes at a rate faster than the denominator, which stabilizes their ratio and ensures smooth convergence to zero.

Overall then, the expression for the correction term has a natural interpretation and exhibits several desirable properties. We emphasize however, that its derivation relies on the assumption that $|\psi_\theta\rangle$ is predominantly supported on the ground state and a single excited state. If this is not true, then Eq.~\eqref{eq:corrected_observable_app} can incur numerical instabilities and may deviate from the targeted ground state observable. Nonetheless, if $|\psi_\theta \rangle$ has appreciable support on the ground state and a few excited states (i.e., more than just one), the above derivation could be generalized to obtain a modified expression for the correction term. Such a generalization would likely require solving a larger system of equations and retaining terms beyond the fourth order nested commutator $M_4(O)$. Extending the our approximation to this setting, and thereby making it a more robust estimator of the ground state observables, would be an interesting direction for future work.

\section{Realizing Approximate Dynamical Averaging in the Heisenberg Picture via Variational Monte Carlo}

In the main text, we applied dynamical averaging to the XXZ model by using the approximation to dynamical averaging in the Heisenberg picture developed in the previous section. By Eq.~\eqref{eq:corrected_observable_app}, applying this to an observable $O$ requires computing the expectation values of the commutators $[H,[H,O]]$ and $[H,[H,[H,[H,O]]]]$. For notational simplicity, let's denote these nested commutators by 
\begin{equation}
    C_n = [H,\cdot]^n(O) , 
\end{equation}
where we use the nested commutator notation defined in Eq.~\eqref{eq:nested_comm_notation}.  We are interested in computing expectation values of $C_2 = [H,[H,O]]$ and $C_4 = [H,[H,[H,[H,O]]]]$.

If $H$ is a local Hamiltonian, and $O$ is a local operator, then these nested commutators will also be sums of local operators. Therefore, one approach to estimating these expectation values is to determine analytically (either by hand or with computer algebra software) expressions for $C_2$ and $C_4$, and then to estimate their expectation values by directly applying variational Monte Carlo to them:
\begin{equation}
    \begin{aligned}
        &\langle \psi_\theta | C_2 |\psi_\theta \rangle = \mathop{\mathbb{E}}_{x \sim |\psi_\theta (x)|^2} \big[ (C_2)_{\text{loc}}(x) \big] \\
        &\langle \psi_\theta | C_4 |\psi_\theta \rangle = \mathop{\mathbb{E}}_{x \sim |\psi_\theta (x)|^2} \big[ (C_4)_{\text{loc}}(x) \big] , 
    \end{aligned}
\end{equation}
where $(C_2)_{\text{loc}}(x) = \frac{\langle x | C_2 |\psi_\theta \rangle}{\langle x | \psi_\theta \rangle}$ and $(C_4)_{\text{loc}}(x) = \frac{\langle x | C_4 |\psi_\theta \rangle}{\langle x | \psi_\theta \rangle}$ are the corresponding local observables.

While this works in principle, these estimators suffer from large variances that require many samples to combat. To reduce variances, we can instead use the following different approach, motivated by the fact that in practice $|\psi_\theta\rangle$ is close to the ground state. To begin, let us use the shift invariance of the commutator to write the expectation value of $C_2$ as 
\begin{equation}
    \begin{aligned}
        &\langle \psi_\theta | C_2 |\psi_\theta \rangle \\
        &= \langle \psi_\theta | [H,C_1] |\psi_\theta \rangle \\
        &= \langle \psi_\theta | [H - E,C_1] |\psi_\theta \rangle \\
        &= \langle \psi_\theta | (H - E)C_1 |\psi_\theta \rangle -\langle \psi_\theta | C_1 (H - E) |\psi_\theta \rangle , 
    \end{aligned}
\end{equation}
for any real constant $E$. Next, note that $C_1 = [H,O]$ is anti-Hermitian ($C_1^\dag = -C_1$), which enables us to express the above equation as  
\begin{equation}\label{eq:C_2_real}
    \begin{aligned}
        & \langle \psi_\theta | (H - E)C_1 |\psi_\theta \rangle -\langle \psi_\theta | C_1 (H - E) |\psi_\theta \rangle \\
        &= \langle \psi_\theta | (H - E)C_1 |\psi_\theta \rangle +\langle \psi_\theta | C_1^\dag (H - E) |\psi_\theta \rangle \\
        &= 2 \cdot \text{Re} \Big[ \langle \psi_\theta | (H - E)C_1 |\psi_\theta \rangle \Big] . 
    \end{aligned}
\end{equation}
The remaining expectation value can then be equated to the following variational Monte Carlo expression by inserting a resolution of the identity in a basis $\{ |x\rangle \} $: 
\begin{equation}\label{eq:C_2_VMC}
    \begin{aligned}
        &\langle \psi_\theta | (H - E) C_1 |\psi_\theta \rangle \\
        &=\sum_x \langle \psi_\theta | (H - E) | x \rangle \langle x| C_1 |\psi_\theta \rangle \\
        &= \sum_x | \psi_\theta(x)|^2 \frac{\langle \psi_\theta | (H - E) | x \rangle}{\langle \psi_\theta | x \rangle } \frac{\langle x| C_1 |\psi_\theta \rangle}{\langle x | \psi_\theta \rangle} \\ 
        &= \mathop{\mathbb{E}}_{x \sim |\psi_\theta (x)|^2} \big[ (H-E)_{\text{loc}}(x) \cdot (C_1)_{\text{loc}}(x) \big] , 
    \end{aligned}
\end{equation}
where $(H-E)_{\text{loc}}(x) = \frac{\langle x | (H - E) | \psi_\theta \rangle}{\langle x | \psi_\theta\rangle }$ and $(C_1)_{\text{loc}}(x) = \frac{\langle x| C_1 |\psi_\theta \rangle}{\langle x | \psi_\theta \rangle}$ are the corresponding local observables. Overall, this provides the expression
\begin{equation}\label{eq:C_2_final}
\begin{aligned}
    &\langle \psi_\theta | C_2 |\psi_\theta \rangle \\
    &= 2 \cdot \text{Re} \Bigg[ \mathop{\mathbb{E}}_{x \sim |\psi_\theta (x)|^2} \big[ (H-E)_{\text{loc}}(x) \cdot (C_1)_{\text{loc}}(x) \big] \Bigg] . 
\end{aligned}
\end{equation}

Lastly, we select $E = \langle \psi_\theta | H | \psi_\theta \rangle$, which reduces the variance of this estimator through the factor $(H-E)_{\text{loc}}(x)$. This occurs because $(H-E)_{\text{loc}}(x)$ approaches $0$ as $|\psi_\theta\rangle$ approaches the ground state. Therefore, with this choice, Eq.~\eqref{eq:C_2_final} dramatically reduces the variance of the estimator for $\langle \psi_\theta | C_2 | \psi_\theta \rangle$, and thus also decreases the number of samples needed to estimate this expectation value accurately. 

Furthermore, we can also derive an analogous expression for $C_4$:
\begin{equation}\label{eq:C_4_final}
    \begin{aligned}
        &\langle \psi_\theta | C_4 |\psi_\theta \rangle \\
        &= 2 \cdot \text{Re} \Big[ \langle \psi_\theta | (H - E)C_1 |\psi_\theta \rangle \Big] \\
        &=2 \cdot \text{Re} \Bigg[ \mathop{\mathbb{E}}_{x \sim |\psi_\theta (x)|^2} \big[ (H-E)_{\text{loc}}(x) \cdot (C_3)_{\text{loc}}(x) \big] \Bigg] . 
    \end{aligned}
\end{equation}
We use Eqs.~\eqref{eq:C_2_final} and ~\eqref{eq:C_4_final} to estimate $\langle \psi_\theta | C_2 |\psi_\theta \rangle $ and $\langle \psi_\theta | C_4 |\psi_\theta \rangle $ in our experiments in the main text.

A further simplification is possible for periodic, translation invariant systems, where the expectation value of a local operator is independent of its position. When the NQS respects this translation symmetry, one may replace a single local operator $O_i$ acting on site $i$ (or more generally, a multi-site operator whose nontrivial support is anchored at site $i$) with its spatial average over the lattice: 
\begin{equation}
    O_{\text{avg}} = \frac{1}{L} \sum_i O_i,
\end{equation}
without changing its expectation value. This spatial averaging provides an additional reduction in variance of the estimator. In our experiments in the main text, we apply this strategy to $C_2$ and $C_4$ to further reduce the number of samples required for accurate estimation of their expectation values.

\section{Details on Numerical Experiments}

In the experiments conducted in the main text, we applied dynamical averaging to various spin models. Here we used both dynamical averaging in the Schr\"odinger picture (i.e., Eq.~\eqref{eq:dyn_avg_eq}), which requires performing real-time evolution on the NQS, as well as our approximation to dynamical averaging in the Heisenberg picture (i.e., Eq.~\eqref{eq:corrected_observable}), which requires computing nested commutators rather than explicit real-time evolution. In both cases, we use NetKet~\cite{netket2:2019, netket3:2022} to implement NQSs, compute observables, perform energy minimization, and perform real-time evolution. 

In applying dynamical averaging in the Schr\"odinger picture, we use different NQS ans\"atze for the 1D and 2D systems. For the 1D system with $N=12$ spins, we use a restricted Boltzmann machine~\cite{Carleo_2017_Solving} with feature density $\alpha=2$, symmetrized under lattice translations, yielding $27$ complex parameters. For the 2D system with $6\times6$ spins, we use a Vision Transformer with translation-invariant factored attention~\cite{viteritti2023transformer,viteritti2025transformer}, consisting of $2\times2$ patches, embedding dimension $d=48$, and two encoder layers with $8$ attention heads and feed-forward expansion factor $2$, yielding $33{,}888$ real parameters. We optimize both ans\"atze using stochastic reconfiguration~\cite{sorella1998green,Sorella_2005}. We use standard stochastic reconfiguration for the 1D system, where $n_{\mathrm{samples}}\gg n_{\mathrm{params}}$, and minimum-step stochastic reconfiguration \cite{chen2024empowering,rende2024simple} for the 2D system, where $n_{\mathrm{params}}\gg n_{\mathrm{samples}}$. The respective learning rates are $\eta=7.5\cdot10^{-4}$ and $10^{-3}$, with $1024$ and $2048$ samples. Both methods use a diagonal regularization shift of $5\cdot10^{-4}$.

To perform dynamical averaging, we choose $p(t)$ as a zero-mean Gaussian with variance $\sigma^2$. We perform real-time evolution using projected time-dependent VMC~\cite{Sinibaldi_2023,netket_fidelity}, which approximates time evolution by minimizing the infidelity between a short-time evolution of the current state and a state within the variational manifold. We use the second-order Pad\'e product expansion~\cite{Gravina_2025}, requiring one compression per time step. For the 1D and 2D systems, respectively, we use time steps $\Delta t=10^{-2}$ and $2.5\cdot10^{-2}$, with $2048$ and $4096$ samples. Starting from selected states along the energy minimization trajectory, we evolve forward and backward over $t\in[-T,T]$, with $T=10$ (1D) and $T=5$ (2D), and evaluate Eq.~\eqref{eq:dyn_avg_eq} with the truncated Gaussian weights renormalized over this interval. To resolve the improvements from dynamical averaging shown in Figs.~\ref{fig:experiments1}a and b, we estimate the observables using additional samples.

In applying our approximation to dynamical averaging in the Heisenberg picture, we also used an NQS parametrized by a restricted Boltzmann machine with feature density $\alpha=2$, symmetrized under lattice translations. For both systems on $N=12$ and $N=40$ spins, we performed energy minimization with the SPRING algorithm ~\cite{Goldshlager_2024}, using learning rate $\eta = 1\cdot10^{-3}$, momentum $\mu=0.95$, and diagonal regularization shift of $1\cdot10^{-3}$. We also used $n_{\text{samples}}=2048$ samples to estimate the gradient of the energy and the quantum geometric tensor during optimization. To compute the expectation values of the nested commutators that comprise Eq.~\eqref{eq:corrected_observable_app}, we used VMC with the estimators derived in the previous section, namely Eqs.~\eqref{eq:C_2_final} and~\eqref{eq:C_4_final}.

\section{High-Accuracy NQSs as an Instance of Randomly-Compiled Quantum Monte Carlo}

Recently, Ref.~\cite{martyn2026faster} developed a framework of \textit{randomly-compiled quantum Monte Carlo}, which enhances quantum Monte Carlo (QMC) algorithms by strategically averaging over approximations. In standard QMC, an observable $\langle O \rangle$ is estimated using a single approximation (e.g., a single Trotter formula). In contrast, randomly-compiled QMC estimates an observable as average over many approximations: 
\begin{equation}\label{eq:RC_QMC}
    \langle O \rangle \approx \sum_j p_j \langle O \rangle_j ,
\end{equation}
where $p_j$ is a probability distribution, and $\langle O \rangle_j$ is the observable obtained using an approximation indexed by $j$. Ref.~\cite{martyn2026faster} showed that this significantly reduces systematic error and provides an improvement over standard QMC algorithms.

Because NQSs incorporate variational Monte Carlo, which is itself a QMC algorithm, they naturally admit a randomly-compiled extension. To wit, this extension would estimate a ground state observable by averaging over a collection ground state approximations. Our dynamical averaging algorithm achieves precisely this. Indeed, Eq.~\eqref{eq:dyn_avg_eq} in the main text proposes to estimate a ground state observable as
\begin{equation}
\begin{aligned}
    &\int dt \ p(t) \langle \psi_\theta | e^{iHt} O e^{-iHt} |\psi_\theta \rangle \\
    &= \int dt \ p(t) \langle \psi_\theta (t) | O | \psi_\theta (t) \rangle . 
\end{aligned}
\end{equation}
Like Eq.~\eqref{eq:RC_QMC}, this expression is manifestly an average of observables evaluated in time-evolved states $|\psi_\theta(t)\rangle$, each of which is an approximation to the ground state. Therefore, dynamical averaging can be interpreted as the randomly-compiled extension of variational Monte Carlo for NQSs.

\end{document}